\documentclass[a4paper,11pt,english]{article}
\pdfoutput=1
\usepackage{jheppub}
\usepackage[utf8]{inputenc}
\usepackage{physics}

\usepackage{tikz}
\usetikzlibrary{patterns}
\usetikzlibrary{decorations.pathmorphing}
\usetikzlibrary{decorations.markings}
\usepgflibrary{arrows.meta}

\pgfdeclarelayer{edgelayer}
\pgfdeclarelayer{nodelayer}
\pgfsetlayers{edgelayer,nodelayer,main}

\usepackage{adjustbox}

\newcommand{\icap}[2] {\texttt{\tiny #1 [#2]}}

\newcommand{\SF}{S}
\newcommand{\IH}{\ensuremath{\mathcal{I}}}
\newcommand{\ep}{\epsilon}

\title{Four-loop HQET propagators from the DRA method}
\author[a]{Roman N. Lee,}
\author[a,b]{Andrey F. Pikelner}
\affiliation[a]{Budker Institute of Nuclear Physics, 630090 Novosibirsk, Russia}
\affiliation[b]{Bogoliubov Laboratory of Theoretical Physics, Joint Institute for Nuclear Research, 141980 Dubna, Russia}
\emailAdd{r.n.lee@inp.nsk.su}
\emailAdd{pikelner@theor.jinr.ru}

\abstract{
  We use dimensional recurrence relations and analyticity to calculate four-loop
  propagator-type master integrals in the heavy-quark effective theory. Compared
  to previous applications of the DRA method, we apply a new technique of fixing
  homogeneous solutions from pole parts of integrals evaluated in different
  rational space-time dimension points. The latter were calculated from the
  integration-by-parts reduction of finite integrals in shifted space-time
  dimension and/or with increased propagators powers. We provide results for epsilon
  expansions of master integrals near $d = 4$ and $d = 3$ using constructed
  alternative sets of integrals with expansion coefficients having conjectural uniform transcendental weight.
}

\begin{document}
\maketitle
\flushbottom

\section{Introduction}
\label{sec:intro}
As is well known, massive internal lines in the diagrams bring much complexity
in the calculations. However, there are two well-known limiting cases where the analysis simplifies: when the lines are massless and they are infinitely heavy.
In particular, this fact justifies the utility of the heavy quark
effective theory(HQET)~\cite{Neubert:1993mb}.
In HQET, in its simplest form, in addition to a single infinitely heavy particle, one considers massless particles only.
Within this theory, the propagator-type integrals are functions with trivial dependence on a single dimensionful parameter, the residual energy $\omega$, and the method of differential equations can not be applied, at least, directly.

Possible applications of such integrals include calculating the heavy quark
field anomalous dimension~\cite{Chetyrkin:2003vi,Grozin:2022wse}, the
small-angle expansion of cusp anomalous dimension, and the correlators of
various currents in HQET~\cite{Chetyrkin:2003vi,Chetyrkin:2021qvd}.
For example, recently, using integrals calculated in the present paper, the four-loop expression for the heavy quark anomalous dimension and first two terms of the small-angle expansion of the QCD cusp anomalous dimension were
calculated in Ref.~\cite{Grozin:2022wse}.

In three-loop calculations there are only eight propagator-type master integrals, and all of them, except one, are known for arbitrary space-time dimension in terms of hypergeometric functions~\cite{Beneke:1994sw,Grozin:2000jv}.
The last non-trivial master integral has been calculated up to $\ep^1$ terms in Ref.~\cite{Czarnecki:2001rh} using its relation to the three-loop on-shell master integral.
Techniques used in three-loop calculation are difficult to apply at four loops due to a large number of master integrals and their grown complexity. Therefore, we choose to switch to a more effective \emph{Dimensional Recurrence and Analyticity} (DRA)
technique~\cite{Lee:2009dh}. This method is based on constructing the general
solutions of the dimensional recurrence relations~\cite{Tarasov:1996br} in the form of triangular series and using the analytical properties of the integrals as the functions of space-time dimension $d$ to fix the undetermined periodic functions.
The derivation of dimensional recurrence relations and the construction of their general solutions can be done rather easily using \texttt{LiteRed}~\cite{Lee:2012cn} and \texttt{SummerTime}~\cite{Lee:2015eva} packages.
In the present paper we demonstrate that the remaining task of fixing the homogeneous solution can also be accomplished in a quasi-automatic fashion.
Our approach is based on deriving required constraints in a specific rational point $d=d_0$ by generating a sufficient number of integrals finite in this point and then reducing them to master integrals.
It appears that the finiteness of the initial integrals provides a highly redundant set of constraints on the expansion coefficients of master integrals around $d_0$, which can be used not only to fix the periodic functions in general solution, but also to safely crosscheck the obtained results.

The paper is organized as follows. In section~\ref{sec:method} we describe some details of the DRA method as applied to our problem, and in section~\ref{sec:Q21-example} we provide an explicit example of the calculation.
Section~\ref{sec:results} contains description of the obtained
results for general $d$ and for $\ep$-expansion near $d=4$ and
$d=3$. We conclude in section~\ref{sec:conclusion}.

\section{Method of calculation}
\label{sec:method}

\tikzstyle{none}=[inner sep=0mm]
\tikzstyle{dot}=[fill=black, draw=black, shape=circle,minimum size=12pt]
\tikzstyle{prop}=[-, draw=red, line width=4pt]
\tikzstyle{legs}=[double distance=6pt, draw=black, line width=2.5pt]
\tikzstyle{HQET}=[double distance=6pt, draw=black, line width=2.5pt]
\begin{figure}
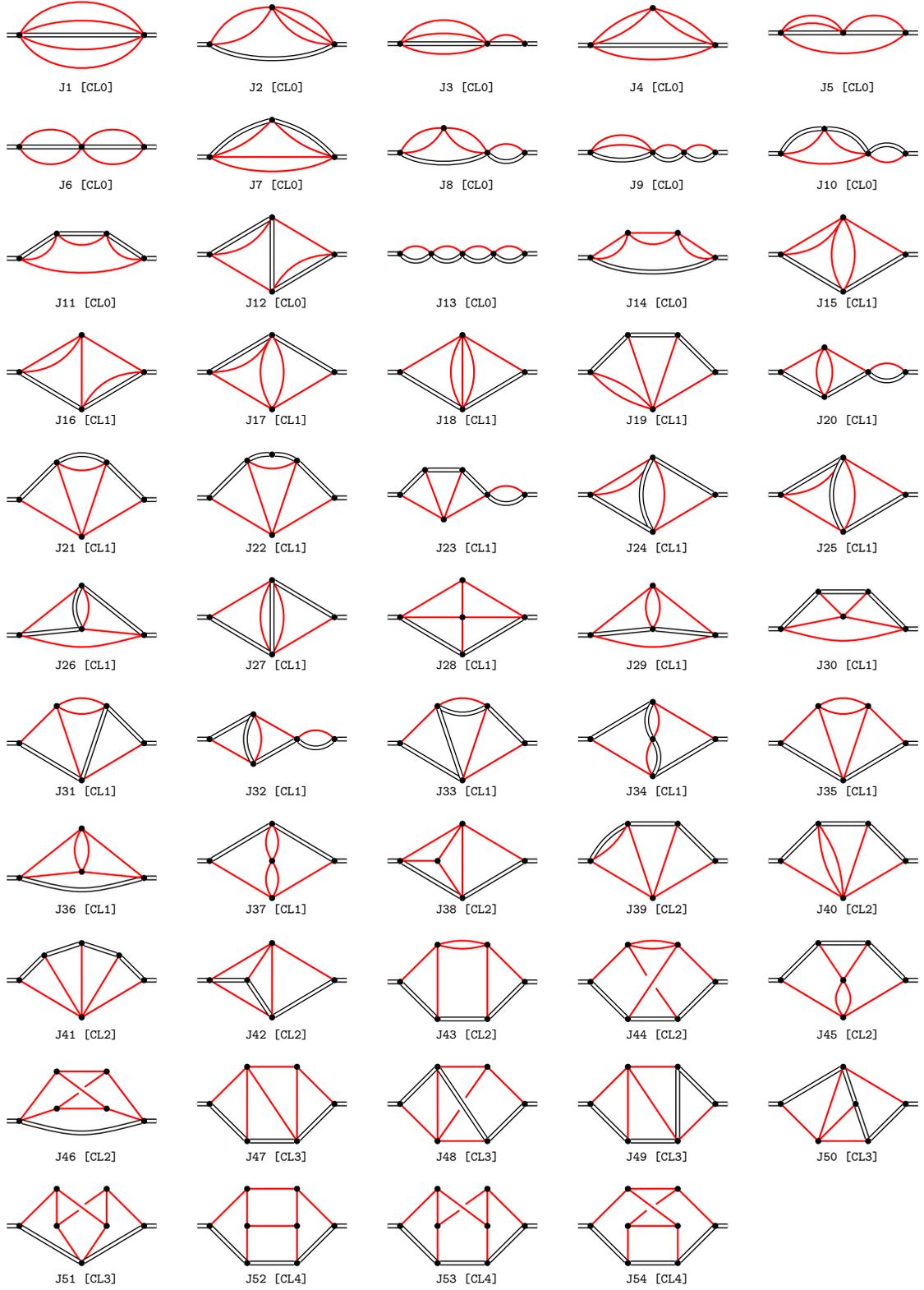

  \centering
  {\renewcommand{\arraystretch}{1.3}
    \begin{tabular}[c]{ccccc}
      
  \adjustbox{valign=c}{\resizebox{2.5cm}{!}{\input figs/CL0/J1.tikz}}

      & 
  \adjustbox{valign=c}{\resizebox{2.5cm}{!}{\input figs/CL0/J2.tikz}}

      & 
  \adjustbox{valign=c}{\resizebox{2.5cm}{!}{\input figs/CL0/J3.tikz}}

      & 
  \adjustbox{valign=c}{\resizebox{2.5cm}{!}{\input figs/CL0/J4.tikz}}

      & 
  \adjustbox{valign=c}{\resizebox{2.5cm}{!}{\input figs/CL0/J5.tikz}}

        \vspace{-8pt}\\
      \icap{J1}{CL0}
      & \icap{J2}{CL0}
      & \icap{J3}{CL0}
      & \icap{J4}{CL0}
      & \icap{J5}{CL0}
        \vspace{6pt}\\
      
  \adjustbox{valign=c}{\resizebox{2.5cm}{!}{\input figs/CL0/J6.tikz}}

      & 
  \adjustbox{valign=c}{\resizebox{2.5cm}{!}{\input figs/CL0/J7.tikz}}

      & 
  \adjustbox{valign=c}{\resizebox{2.5cm}{!}{\input figs/CL0/J8.tikz}}

      & 
  \adjustbox{valign=c}{\resizebox{2.5cm}{!}{\input figs/CL0/J9.tikz}}

      & 
  \adjustbox{valign=c}{\resizebox{2.5cm}{!}{\input figs/CL0/J10.tikz}}

        \vspace{-8pt}\\
      \icap{J6}{CL0}
      & \icap{J7}{CL0}
      & \icap{J8}{CL0}
      & \icap{J9}{CL0}
      & \icap{J10}{CL0}
        \vspace{6pt}\\
      
  \adjustbox{valign=c}{\resizebox{2.5cm}{!}{\input figs/CL0/J11.tikz}}

      & 
  \adjustbox{valign=c}{\resizebox{2.5cm}{!}{\input figs/CL0/J12.tikz}}

      & 
  \adjustbox{valign=c}{\resizebox{2.5cm}{!}{\input figs/CL0/J13.tikz}}

      & 
  \adjustbox{valign=c}{\resizebox{2.5cm}{!}{\input figs/CL0/J14.tikz}}

      & 
  \adjustbox{valign=c}{\resizebox{2.5cm}{!}{\input figs/CL1/J15.tikz}}

        \vspace{-8pt}\\
      \icap{J11}{CL0}
      & \icap{J12}{CL0}
      & \icap{J13}{CL0}
      & \icap{J14}{CL0}
      & \icap{J15}{CL1}
        \vspace{6pt}\\
      
  \adjustbox{valign=c}{\resizebox{2.5cm}{!}{\input figs/CL1/J16.tikz}}

      & 
  \adjustbox{valign=c}{\resizebox{2.5cm}{!}{\input figs/CL1/J17.tikz}}

      & 
  \adjustbox{valign=c}{\resizebox{2.5cm}{!}{\input figs/CL1/J18.tikz}}

      & 
  \adjustbox{valign=c}{\resizebox{2.5cm}{!}{\input figs/CL1/J19.tikz}}

      & 
  \adjustbox{valign=c}{\resizebox{2.5cm}{!}{\input figs/CL1/J20.tikz}}

        \vspace{-8pt}\\
      \icap{J16}{CL1}
      & \icap{J17}{CL1}
      & \icap{J18}{CL1}
      & \icap{J19}{CL1}
      & \icap{J20}{CL1}
        \vspace{6pt}\\
      
  \adjustbox{valign=c}{\resizebox{2.5cm}{!}{\input figs/CL1/J21.tikz}}

      & 
  \adjustbox{valign=c}{\resizebox{2.5cm}{!}{\input figs/CL1/J22.tikz}}

      & 
  \adjustbox{valign=c}{\resizebox{2.5cm}{!}{\input figs/CL1/J23.tikz}}

      & 
  \adjustbox{valign=c}{\resizebox{2.5cm}{!}{\input figs/CL1/J24.tikz}}

      & 
  \adjustbox{valign=c}{\resizebox{2.5cm}{!}{\input figs/CL1/J25.tikz}}

        \vspace{-8pt}\\
      \icap{J21}{CL1}
      & \icap{J22}{CL1}
      & \icap{J23}{CL1}
      & \icap{J24}{CL1}
      & \icap{J25}{CL1}
        \vspace{6pt}\\
      
  \adjustbox{valign=c}{\resizebox{2.5cm}{!}{\input figs/CL1/J26.tikz}}

      & 
  \adjustbox{valign=c}{\resizebox{2.5cm}{!}{\input figs/CL1/J27.tikz}}

      & 
  \adjustbox{valign=c}{\resizebox{2.5cm}{!}{\input figs/CL1/J28.tikz}}

      & 
  \adjustbox{valign=c}{\resizebox{2.5cm}{!}{\input figs/CL1/J29.tikz}}

      & 
  \adjustbox{valign=c}{\resizebox{2.5cm}{!}{\input figs/CL1/J30.tikz}}

        \vspace{-8pt}\\
      \icap{J26}{CL1}
      & \icap{J27}{CL1}
      & \icap{J28}{CL1}
      & \icap{J29}{CL1}
      & \icap{J30}{CL1}
        \vspace{6pt}\\
      
  \adjustbox{valign=c}{\resizebox{2.5cm}{!}{\input figs/CL1/J31.tikz}}

      & 
  \adjustbox{valign=c}{\resizebox{2.5cm}{!}{\input figs/CL1/J32.tikz}}

      & 
  \adjustbox{valign=c}{\resizebox{2.5cm}{!}{\input figs/CL1/J33.tikz}}

      & 
  \adjustbox{valign=c}{\resizebox{2.5cm}{!}{\input figs/CL1/J34.tikz}}

      & 
  \adjustbox{valign=c}{\resizebox{2.5cm}{!}{\input figs/CL1/J35.tikz}}

        \vspace{-8pt}\\
      \icap{J31}{CL1}
      & \icap{J32}{CL1}
      & \icap{J33}{CL1}
      & \icap{J34}{CL1}
      & \icap{J35}{CL1}
        \vspace{6pt}\\
      
  \adjustbox{valign=c}{\resizebox{2.5cm}{!}{\input figs/CL1/J36.tikz}}

      & 
  \adjustbox{valign=c}{\resizebox{2.5cm}{!}{\input figs/CL1/J37.tikz}}

      & 
  \adjustbox{valign=c}{\resizebox{2.5cm}{!}{\input figs/CL2/J38.tikz}}

      & 
  \adjustbox{valign=c}{\resizebox{2.5cm}{!}{\input figs/CL2/J39.tikz}}

      & 
  \adjustbox{valign=c}{\resizebox{2.5cm}{!}{\input figs/CL2/J40.tikz}}

        \vspace{-8pt}\\
      \icap{J36}{CL1}
      & \icap{J37}{CL1}
      & \icap{J38}{CL2}
      & \icap{J39}{CL2}
      & \icap{J40}{CL2}
        \vspace{6pt}\\
      
  \adjustbox{valign=c}{\resizebox{2.5cm}{!}{\input figs/CL2/J41.tikz}}

      & 
  \adjustbox{valign=c}{\resizebox{2.5cm}{!}{\input figs/CL2/J42.tikz}}

      & 
  \adjustbox{valign=c}{\resizebox{2.5cm}{!}{\input figs/CL2/J43.tikz}}

      & 
  \adjustbox{valign=c}{\resizebox{2.5cm}{!}{\input figs/CL2/J44.tikz}}

      & 
  \adjustbox{valign=c}{\resizebox{2.5cm}{!}{\input figs/CL2/J45.tikz}}

        \vspace{-6pt}\\
      \icap{J41}{CL2}
      & \icap{J42}{CL2}
      & \icap{J43}{CL2}
      & \icap{J44}{CL2}
      & \icap{J45}{CL2}
        \vspace{6pt}\\
      
  \adjustbox{valign=c}{\resizebox{2.5cm}{!}{\input figs/CL2/J46.tikz}}

      & 
  \adjustbox{valign=c}{\resizebox{2.5cm}{!}{\input figs/CL3/J47.tikz}}

      & 
  \adjustbox{valign=c}{\resizebox{2.5cm}{!}{\input figs/CL3/J48.tikz}}

      & 
  \adjustbox{valign=c}{\resizebox{2.5cm}{!}{\input figs/CL3/J49.tikz}}

      & 
  \adjustbox{valign=c}{\resizebox{2.5cm}{!}{\input figs/CL3/J50.tikz}}

        \vspace{-6pt}\\
      \icap{J46}{CL2}
      & \icap{J47}{CL3}
      & \icap{J48}{CL3}
      & \icap{J49}{CL3}
      & \icap{J50}{CL3}
        \vspace{6pt}\\
      
  \adjustbox{valign=c}{\resizebox{2.5cm}{!}{\input figs/CL3/J51.tikz}}

      & 
  \adjustbox{valign=c}{\resizebox{2.5cm}{!}{\input figs/CL4/J52.tikz}}

      & 
  \adjustbox{valign=c}{\resizebox{2.5cm}{!}{\input figs/CL4/J53.tikz}}

      & 
  \adjustbox{valign=c}{\resizebox{2.5cm}{!}{\input figs/CL4/J54.tikz}}

      &
        \vspace{-6pt}\\
      \icap{J51}{CL3}
      & \icap{J52}{CL4}
      & \icap{J53}{CL4}
      & \icap{J54}{CL4}
      &
    \end{tabular}}
  \caption{Four-loop propagator-type HQET integrals calculated in this paper.
    Red solid lines correspond to massless propagators $[-l^2-i0]^{-1}$, double
    lines correspond to the HQET propagator $[1-2l\cdot n-i0]^{-1}$, (where
    $n^2=1$). The loop integration measure is $\frac{d^dl}{i\pi^{d/2}}$.}
  \label{fig:MIs}
\end{figure}
We consider the propagator-type diagrams obtained by attaching an $n$-legged
graph with massless lines to $n$ points on a single HQET line. For four loops we
have $2\leqslant n\leqslant 8$, however for the master integrals identification
we can restrict ourselves with $n=3,\,4$, or $5$ as other cases reduce to these
by partial fractioning. Performing the IBP reduction of the remaining 19 big
topologies, we end up with 54 master integrals shown in Fig.\ref{fig:MIs}. We
also find the dimensional recurrence relations of the form
\begin{equation}
  \label{eq:LDRR}
	\boldsymbol{J} (d+2)=L(d)
	\boldsymbol{J} (d)
\end{equation}
The application of the DRA method is straightforward provided the matrix $L(d)$
is lower-triangular. This property is obvious for the case when there are no
more than one master integral in each sector. In our case the
integrals $J_{21}$ and $J_{22}$ belong to the same sector, but,fortunately, the corresponding block in the matrix $L$ is diagonal.
The triangular form of the matrix $L(d)$ results to the first-order inhomogeneous difference equation for each $J_k$:
\begin{equation}\label{eq:Q_k:DRR}
	J_k(d+2)=L_{kk}(d) J_k(d)+\sum_{l<k} L_{kl}(d) J_l(d).
\end{equation}
Let us assume that $J_l(d)$ for $l<k$ are already calculated by the same method.
Then the general solution of Eq. \eqref{eq:Q_k:DRR} can be written as
\begin{equation}\label{eq:Q_k:Sol}
  J_k(d)=\SF^{-1}(d)\omega(d) + \mathcal{R}_k (d)\,,
\end{equation}
where $\mathcal{R}_k (d)$ is a specific solution of inhomogeneous equation,
$\omega(d)=\omega(d+2)$ is an arbitrary periodic factor, and the \textit{summing
factor} $\SF(d)$ is a specific solution of
\begin{equation}\label{eq:DRRhom}
  \SF(d)=L_{kk}(d)\SF(d+2)\,.
\end{equation}
The specific inhomogeneous solution $\mathcal{R}_k$ can be expressed in terms of
triangular sums, see Ref. \cite{Lee:2009dh} for details.

In order to fix the function $\omega(z)$ we need to obtain sufficient
information about the analytical properties of $J_k$ as a function of $d$ on the
chosen \textit{basic stripe} $\mathcal{B}$ --- a vertical stripe of width $2$ in
the complex plane of $d$. We find it sufficient to use for all master integrals
$J_k$ one and the same basic stripe
\begin{equation}
  \mathcal{B}=\{d\in\mathbb{C}|\ 0<\Re d\leqslant 2\}\,.
\end{equation}
The conventional approach to obtain the required analytical data is the
following. First, one defines the positions of possible singularities on the
basic stripe using \texttt{Fiesta}'s \cite{Smirnov2016} routine
\texttt{SDAnalyze}. Then, for each of the found position one tries to find the
order of the pole and a few leading coefficients of Laurent expansion near it.
For these goals one often needs to derive the Mellin-Barnes representation and
to use the \texttt{MB} code \cite{Czakon2006}, although for some simple cases it
might be sufficient to use \texttt{Fiesta}'s routine \texttt{SDEvaluate} alone.
This approach usually gives a few first terms of Laurent expansion as
limited-precision numbers and should be complemented by the educated guess about
their analytical form.

The main drawback of this approach is that the derivation of the required
Mellin-Barnes representation requires a substantial amount of manual work. In
the present paper we use an alternative approach which appears to provide more
than enough information about the analytical properties of the master integrals.
First, instead of using \texttt{Fiesta}'s \texttt{SDAnalyze}, we follow a less
specific but much more simple way to restrict possible positions of
singularities. Namely, we analyze the position of poles in the matrix of
dimensional recurrence relations and assume that the singularities may differ
from those poles by a multiple of $2$. In this way we obtain the following set
of potential positions of poles on the basic stripe:
\begin{equation}
  \mathcal{S}=\left\{\tfrac{1}{4},\tfrac{1}{3},\tfrac{2}{5},\tfrac{1}{2},\tfrac{2}{3},\tfrac{3}{4},\tfrac{4}{5},1,\tfrac{6}{5},\tfrac{5}{4},\tfrac{4}{3},\tfrac{3}{2},\tfrac{8}{5},\tfrac{5}{3},\tfrac{7}{4},2\right\}.
\end{equation}
The main step of our approach is to consider large enough set of integrals in
some chosen $d=d_0-2\ep+2k$, with $d_0\in \mathcal{S}$ and $k\in \mathbb{Z}$,
\emph{finite} at $\ep=0$ and to reduce them to master integrals
$J_1,\ldots,J_{54}$ in $d=d_0-2\ep$ using IBP identities and dimensional
recurrence relations. Then, the finiteness of the obtained expression implies
some constraints on the $\ep$-expansion coefficients of master integrals at
$d=d_0-2\ep$. This approach is very similar to the one that was successfully
used for the calculation of four-loop~\cite{Baikov:2010hf} and
five-loop~\cite{Georgoudis:2021onj} massless propagators\footnote{In the
calculations of massless propagators it was important to use also the additional
Glue-and-Cut symmetry.}. The only difference is that for our present purposes we
need to consider not a single value of $d_0$, but all points in $\mathcal{S}$.
In order to pick a set of finite integrals we use the algorithm of Ref.
\cite{vonManteuffel:2014qoa} as implemented in the public code \texttt{Reduze2}
\cite{vonManteuffel:2012np}. As the existing implementation supports only even
$d_0$, we had to slightly modify the routines of \texttt{Reduze2} code to
support arbitrary rational $d_0$.

\section{Example: \boldmath{$J_{21}$} integral}
\label{sec:Q21-example}

Let us describe our method in some details on the example of integral $J_{21}$,
\begin{equation}\label{eq:Q21def}
	J_{21}(d)=\int\frac{dl_1dl_2dl_3dl_4/(i\pi^{d/2})^4}{[-l_1^2] [-l_{13}^2] [-l_{23}^2] [-l_{34}^2] [-l_4^2] [1-2 l_1\cdot n] [1-2 l_2\cdot n][1-2 l_4\cdot n]}\,
\end{equation}
where $n$ is a unit time-like vector, $n^2=1$, $l_{ik}=l_i-l_k$, and $[a]=(a-i0)$.
The integral $J_{21}$ satisfies the equation
\begin{equation}\label{eq:DRR21}
  J_{21}(d+2)=c_{21}(d)J_{21}(d)+c_{7}(d)J_{7}(d)+c_{4}(d)J_{4}(d)+c_{3}(d)J_{3}(d)\,,
\end{equation}
where $c_k(d)$ are some rational functions of $d$, in particular,
\begin{equation}\label{eq:c21}
  c_{21}(d)=-\frac{3 (d-3) (3 d-7) (3 d-5)}{16 (d-1)^3 (4 d-11) (4 d-9) (4 d-7) (4 d-5)}
\end{equation}

\paragraph{Analytical properties.} In order to discover the analytical
properties of $J_{21}$, we determine the set of finite integrals for each point
in $\mathcal{S}$. For example, we find that the integral
$\widetilde{J}_{21}=\vcenter{\hbox{\resizebox{1.5cm}{!}{\begin{tikzpicture}
	\begin{pgfonlayer}{nodelayer}
		\node [style=none] (0) at (-5, 0) {};
		\node [style=none] (1) at (5, 0) {};
		\node [style=none] (4) at (6, 0) {};
		\node [style=none] (5) at (-6, 0) {};
		\node [style=dot] (14) at (-5, 0) {};
		\node [style=dot] (15) at (5, 0) {};
		\node [style=dot] (16) at (3, 2) {};
		\node [style=dot] (17) at (-3, 2) {};
		\node [style=dot] (18) at (0, -3) {};
		\node [style=dot] (19) at (-2, 2.5) {};
		\node [style=dot] (20) at (-0.75, 2.75) {};
		\node [style=dot] (21) at (0.75, 2.75) {};
		\node [style=dot] (22) at (2, 2.5) {};
	\end{pgfonlayer}
	\begin{pgfonlayer}{edgelayer}
		\draw [style=legs] (1.center) to (4.center);
		\draw [style=legs] (5.center) to (0.center);
		\draw [style=HQET] (14) to (17);
		\draw [style=prop, bend right] (17) to (16);
		\draw [style=HQET] (16) to (15);
		\draw [style=prop] (14) to (18);
		\draw [style=prop] (18) to (15);
		\draw [style=prop] (17) to (18);
		\draw [style=prop] (16) to (18);
		\draw [style=HQET] (17) to (19);
		\draw [style=HQET] (19) to (20);
		\draw [style=HQET] (20) to (21);
		\draw [style=HQET] (21) to (22);
		\draw [style=HQET] (22) to (16);
	\end{pgfonlayer}
\end{tikzpicture}
}}}$
is finite in $d=4$. Reducing this integral in $d=4-2\ep$ to the master integrals
in $d=2-2\ep$, we obtain
\begin{multline}
  \widetilde{J}_{21}(4-2\ep)=\frac{\ep  (1+2 \ep)^2J_{21}(2-2\ep)}{16 (1-2 \ep) (1+8 \ep) (3+8 \ep)}
  +\frac{ \ep  \left(386 \ep ^3+395 \ep ^2+131 \ep +14\right)J_7(2-2\ep)}{48 (1-2 \ep) (1+3 \ep) (2+5 \ep) (1+8 \ep)}\\
  +\frac{\ep  \left(53 \ep ^2+37 \ep +6\right) (2 \ep +1)J_4(2-2\ep) }{64 (2 \ep -1) (4 \ep +1) (5 \ep +2) (8 \ep +1)}
  +\frac{\ep  (6 \ep +1) \left(124 \ep ^2+95 \ep +18\right)J_3(2-2\ep)}{48 (2 \ep -1) (5 \ep +2) (8 \ep +1) (8 \ep +3)}
\end{multline}
Since the left-hand side is finite, so is the right-hand side.
Note that in the latter the integral $J_{21}$ is the only nontrivial one, while $J_3,\, J_4,\, J_7$ are expressed in terms of $\Gamma$-functions.
Expanding the right-hand side up to $\ep^{-1}$, we obtain the following constraint
\begin{equation}
  e^{4\ep \gamma_E}J_{21}(2-2\ep) =
  -\frac{10}{\ep ^4}-\frac{226}{3 \ep ^3}+\left(\frac{286}{3}-58 \pi ^2\right)\ep ^{-2}+\mathcal{O}(\ep^{-1})
\end{equation}
In fact, we can obtain yet more terms of expansion of $J_{21}$ near $d=2$ once we consider more finite integrals.
Finally, we obtain
\begin{align}
  \begin{split}
    e^{4\ep \gamma_E}J_{21}(2-2\ep) =
    -\frac{10}{\ep ^4}-\frac{226}{3 \ep ^3}+\left(\frac{286}{3}-58 \pi ^2\right)\ep ^{-2} +\left(\frac{5512 \zeta (3)}{3}-\frac{166}{3}-\frac{3826 \pi ^2}{9}\right)\ep^{-1}\\
    + \frac{118336 \zeta (3)}{9}-\frac{728}{3}+\frac{4904 \pi ^2}{9}-\frac{4478 \pi ^4}{15}+\mathcal{O}\left(\ep\right)\,,
  \end{split}\nonumber\\
  \begin{split}
    e^{4\ep \gamma_E}J_{21}(1-2\ep)=
    -3072 \pi ^2+ \left(-\frac{1084928 \pi ^2}{45}-24576 \pi ^2 \log {2}\right)\ep+\mathcal{O}\left(\ep ^2\right)\,,
  \end{split}\nonumber\\
  \begin{split}
    e^{4\ep \gamma_E}J_{21}(2/3-2\ep)=
    -\frac{14554000 \Gamma \left(\frac{4}{3}\right)^5}{189 \ep }+\mathcal{O}\left(\ep ^0\right)\,,
  \end{split}\nonumber\\
  \begin{split}
    e^{4\ep \gamma_E}J_{21}(4/3-2\ep)=
    \frac{16677 \Gamma \left(\frac{5}{3}\right)^5}{10 \ep }+\mathcal{O}\left(\ep ^0\right)\,,
  \end{split}\nonumber\\
  \begin{split}
    e^{4\ep \gamma_E}J_{21}(d_0-2\ep)= \mathcal{O}\left(\ep ^0\right)\text{ in all other points } d_0\in (0,2]\,.
  \end{split}
  \label{eq:constraint}
\end{align}
The right-hand sides of these constraints are built from the expansion
coefficients of simpler integrals $J_{1,2,3,5}$ expressible in terms of
$\Gamma$-functions.

\paragraph{Summing factor.} The summing factor $\SF(d)$ satisfies
\begin{equation}\label{eq:SigmaEq}
  \SF_{i}^{-1}(d+2)=c_{i}(d)\SF_{i}^{-1}(d)\,,
\end{equation}
where $c_{21}$ is defined in Eq. \eqref{eq:c21}.
It is useful to consider its following decomposition
\begin{equation}\label{eq:Sigma}
	\SF(d) = \SF_0(d) \Omega(d)f(d)
\end{equation}
where each of the factors $\SF_0(d),\ \Omega(z),\ f$ has its own meaning.
First, we find $\SF_0(d)$, which is a random solution of Eq. \eqref{eq:SigmaEq}.
Then we pick a periodic factor $\Omega(z)=\Omega(e^{i\pi d})$ such as to reduce the number and the orders of singularities of the quantity  $\SF(d)J_{21}(d)$ on the basic stripe.
In addition we secure that $\SF(d)J_{21}(d)/e^{\pi |d|}$ decays when $d\to\pm i\infty$.
Finally, we pick a constant factors $f$ to simplify the leading term of expansion of $\SF(d)$ at $d=2-2\ep$.
We have
\begin{align}%
  \SF_0(d) & =\frac{2^{4 d}  \Gamma \left(\frac{11}{2}-\frac{3 d}{2}\right)}{ \Gamma \left(\frac{13}{2}-2 d\right) \Gamma \left(\frac{3}{2}-\frac{d}{2}\right)^3}\\
  \Omega(z) & = \sin ^3\left(\tfrac{\pi}{2}   (d-2)\right) \sin \left(\tfrac{ \pi }{2} \left(d-\tfrac{4}{3}\right)\right) \sin \left(\tfrac{\pi}{2}   \left(d-\tfrac{2}{3}\right)\right)\\
  f & = \frac{1}{192 \pi ^{3/2}}\label{eq:f}
\end{align}

Consequently, we have the following properties of $\SF(d)$:
\begin{enumerate}
\item $\SF(d)$ satisfies Eq. \eqref{eq:SigmaEq}.
\item $\SF(d)J_{21}(d)$ has no singularities at $d\in (0,2)$ and is bounded when $\Im d\to\pm \infty$.
\item $\SF(2-2\ep)J_{21}(2-2\ep)=\frac{10}{\ep }+\frac{146}{3}+\mathcal{O}\left(\ep \right)$.
\end{enumerate}
Two last properties are trivially established from the constraints \eqref{eq:constraint}.

\paragraph{General and specific solution.} We write the general solution of Eq. \eqref{eq:DRR21} as
\begin{gather}\label{eq:gensol}
  \SF(d)J_{21}(d)=\IH_{21}(d)+\omega(z)\,,\\
  \label{eq:IH}
  \IH_{21}(d)=-\sum_{k=0}^{\infty}\SF(d_k+2)\left[(c_{7}(d_k)J_{7}(d_k)+c_{4}(d_k)J_{4}(d_k)+c_{3}(d_k)J_{3}(d_k)\right]
\end{gather}
where $d_k=d+2k$ and $\omega(z)=\omega(e^{i\pi d})$ is a periodic function.

Now we have to construct $\omega(z)$ in the right-hand side to fit the analytical properties of the left-hand side of Eq. \eqref{eq:gensol}.
We use \texttt{SummerTime} package \cite{Lee:2015eva} to calculate with high numerical precision the coefficients of $\ep$-expansion of the inhomogeneous solution $\IH_{21}(d)$ near points $d_0\in \mathcal{S}$.
Then, using some educated guess about the transcendental constants which may appear in the coefficients, we obtain the following analytic expansions
\begin{align}
  \IH_{21}(1-2\ep) & \stackrel{\texttt{PSLQ}}{=} \tfrac{1}{27\ep}+\mathcal{O}(\ep^0),
  & \IH_{21}(\tfrac12-2\ep) & \stackrel{\texttt{PSLQ}}{=} \tfrac{1}{2\ep}+\mathcal{O}(\ep^0),\nonumber\\
  \IH_{21}(\tfrac32-2\ep) & \stackrel{\texttt{PSLQ}}{=} \tfrac{1}{2\ep}+\mathcal{O}(\ep^0),
  & \IH_{21}(\tfrac13-2\ep) & \stackrel{\texttt{PSLQ}}{=} \tfrac{1}{9 \sqrt{3} \pi  \epsilon ^2}-\tfrac{14}{27 \epsilon }+\mathcal{O}(\ep^0), \nonumber\\
  \IH_{21}(\tfrac53-2\ep) & \stackrel{\texttt{PSLQ}}{=} -\tfrac{1}{9 \sqrt{3} \pi  \epsilon ^2}-\tfrac{14}{27 \epsilon }+\mathcal{O}(\ep^0),
  & \IH_{21}(2-2\ep) & \stackrel{\texttt{PSLQ}}{=} \tfrac{10}{\epsilon }+\tfrac{146}{3}+\mathcal{O}\left(\ep^1\right)\,,
\end{align}
and $\IH_{21}(d_0-2\ep)=\mathcal{O}(\ep^0)$ for all other $d_0$ from $\mathcal{S}$.
Here $\stackrel{\texttt{PSLQ}}{=}$ means that the analytic coefficients in $\ep$-expansion have been determined from their multi-digit numerical values using \texttt{PSLQ}, Ref. \cite{FergBai1991}.
Then we construct the function $\omega(z)$ which cancels the poles at $d=1,\tfrac12,\tfrac32,\tfrac13,\tfrac53$ and preserves the expansion at $d=2$ and the behavior of $\SF(d)J_{21}(d)$ at $d\to\pm i\infty$:
\begin{multline}\label{eq:omega}
  \omega(z) \stackrel{\texttt{PSLQ}}{=} \tfrac{\pi }{9 \sqrt{3}} \cot ^2{\tfrac{\pi}{2}  \left(d-\tfrac{5}{3}\right)}
  -\tfrac{14 \pi}{27}   \cot {\tfrac{\pi}{2}  \left(d-\tfrac{5}{3}\right)}
  -\tfrac{\pi}{9 \sqrt{3}}\cot ^2{\tfrac{\pi}{2}  \left(d-\tfrac{1}{3}\right)}\\
  -\tfrac{14 \pi}{27}   \cot {\tfrac{\pi}{2}  \left(d-\tfrac{1}{3}\right)}
  + \tfrac{\pi}{2}  \cot {\tfrac{\pi}{2}  \left(d-\tfrac{3}{2}\right)}
  + \tfrac{\pi}{27}   \cot {\tfrac{\pi}{2}  (d-1)}
  + \tfrac{\pi}{2}  \cot {\tfrac{\pi}{2}  \left(d-\tfrac{1}{2}\right)} \\
  = -\frac{2 \pi  \sin \left(\frac{\pi  d}{2}\right) (1-2 \cos (2 \pi  d))}{3 (1-2 \cos (\pi  d))^2 \left(\cos \left(\frac{\pi  d}{2}\right)+\cos \left(\frac{3 \pi  d}{2}\right)\right)}\,,
\end{multline}
Thus we obtain the final expression
\begin{equation}
  J_{21}(d)=S^{-1}(d)\left[\IH_{21}(d)+\omega(z)\right]\,,
\end{equation}
where $S(d)$, $\IH_{21}(d)$, and $\omega(d)$ are defined in Eqs. \eqref{eq:Sigma}--\eqref{eq:f}, \eqref{eq:IH}, and \eqref{eq:omega}, respectively.
Using the \texttt{SummerTime} package to calculate the sum in $\IH_{21}(d)$, we obtain the $\ep$ expansion around $d=4$ and $d=3$ with high-precision numeric coefficients.
Using \texttt{PSLQ}, we obtain
\begin{align}
	J_{21}(4-2\ep)&=e^{-4\ep \gamma_E}\bigg[
                  -\left(\tfrac{7}{288} + \tfrac{\zeta_2}{36}\right)\tfrac{1}{\ep^2}
                  - \left(\tfrac{667}{1728} + \tfrac{8 \zeta_2}{27} - \tfrac{5 \zeta_3}{72}\right)\tfrac{1}{\ep}
                  -\tfrac{31993}{10368}
                  - \tfrac{2725 \zeta_2}{1296} + \tfrac{20 \zeta_3}{27} - \tfrac{49 \zeta_2^2}{360}\nonumber\\
                &\qquad\qquad
                  - \left(\tfrac{636223}{62208} + \tfrac{79321 \zeta_2}{7776} + \tfrac{196 \zeta_2^2}{135}
                  - \tfrac{835\zeta_3}{324} + \tfrac{293 \zeta_2 \zeta_3}{108}
                  + \tfrac{191\zeta_5}{24} \right) \ep + \ldots\bigg]\\
	J_{21}(3-2\ep)&=e^{-4\ep \gamma_E} \pi^2\bigg[
                  \tfrac{7  \zeta _3}{\ep }+\tfrac{14}{5} \zeta _2^2+32 \zeta_{-3,1}
                  +\left(254 \zeta _2 \zeta _3+357 \zeta_5-256 \zeta_{-3,1,1}\right) \ep
                  \nonumber\\
                &\qquad \qquad
                  + \left(\tfrac{5172}{35} \zeta _2^3+192 \zeta _{-3,1} \zeta_2 - \tfrac{2746}{3} \zeta _3^2
                  - 640 \zeta _{-5,1}+2048 \zeta_{-3,1,1,1}\right) \ep^2+\ldots
                  \bigg]\,,
\end{align}
where $\zeta_{a_1\dots a_n}$ is defined in~(\ref{eq:mzv-def}).

\section{Results}
\label{sec:results}

Similar to the example in previous section, we derive representation for all
master integrals from Fig.~\ref{fig:MIs} in terms of iterated triangular sums
with factorized summands. One can effectively evaluate these sums as expansions
in $\ep$ with arbitrarily accurate numerical coefficients using the
\texttt{SummerTime} package \cite{Lee:2015eva}. Assuming that we know the basis
of transcendental numbers, which may show up in the results, we can use the
\texttt{PSLQ} algorithm to recover the analytical form of the coefficients.

With the paper, we provide the results for sums in the \texttt{SummerTime}
format, admitting the calculation of all considered integrals for arbitrary
space-time dimension and/or to arbitrary order in $\ep$. Furthermore, we perform
\texttt{PSLQ} recognition for $d=3-2\ep$ and $d=4-2\ep$ to obtain the analytic
results, which should be sufficient for any practical application. For
$d=4-2\ep$ we successfully recognize the analytic result in terms of usual
multiple zeta values. For $d=3-2\ep$ we use also the alternating Euler-Zagier
sums.

The existence of a uniformly transcendental (UT) basis is very remarkable per
se, but it is advantageous also for practical reasons as it simplifies the
\texttt{PSLQ} recognition. Unfortunately, it is yet unclear how to
systematically construct the UT basis for one-scale integrals. Nevertheless, we
were able to construct UT bases for both three-dimensional and four-dimensional
cases. This was accomplished in a semi-empirical way by checking the integrals
which diverge logarithmically in $d=4$ or $d=3$. In many cases we observed that
such integrals exhibit the property of uniform transcendentality after pulling
out a simple rational factor. We have recognized the analytical results for UT
integrals up to the weight twelve\footnote{Note that the expansion of integrals
at $d=3-2\ep$ appears to have an overall common factor $\pi^2$, so after pooling
this factor out we had to recognize only up to t.w. 10 expressions.} for both
$d=4-2\ep$ and $d=3-2\ep$.

The obtained results are attached to the \texttt{arXiv} version of the paper.
The description of the attached files can be found in Appendix
~\ref{sec:anc-files}.

\paragraph{Checks of the results.}
Since our method is quite involved, it is important to perform some crosschecks
of our results. First, as we mentioned earlier, the IBP reduction of finite
integrals provide an extremely redundant set of constraints which we use not
only for fixing the specific solution of dimensional recurrence relations, but
also for an extensive crosscheck of the results. Then, we have verified that our
results reproduce all terms of $\ep$ expansion of integrals calculated in
\cite{Grozin:2017css} for $d=4-2\ep$. Unfortunately, for the most complicated
integrals in that paper only the divergent part is available. While this paper
was being written a work on the numerical calculation of the same set of
four-loop integrals appeared~\cite{Liu:2022tji}. The comparison of the results
provided therein with those of the present paper for the cases $d=4-2\ep$ has
shown only partial agreement\footnote{In our notations we found disagreement in
  integrals $J_{28}$, $J_{30}$, $J_{36}$, $J_{44}$, $J_{45}$, $J_{46}$, $J_{52}$, $J_{53}$, $J_{54}$.}. We note that the results of Ref.~\cite{Liu:2022tji} for
the most complicated integrals $J_{52}$, $J_{53}$, $J_{54}$ are also in
contradiction with the divergent parts of these integrals calculated
in~\cite{Grozin:2017css}.

\section{Conclusion}
\label{sec:conclusion}
In this paper we have calculated the four-loop HQET propagator-type master
integrals using the DRA method \cite{Lee:2009dh}. In order to fix the periodic
functions in the general solution of the dimensional recurrence relations, we
use a novel highly automated approach based on the IBP reduction of finite
integrals. In order to pick a sufficient set of the finite integrals we use a
criterion from \cite{vonManteuffel:2014qoa} generalized to the case of rational
$d$. Having obtained the expressions for the master integrals in terms of
triangular sums treatable by the \texttt{SummerTime} package \cite{Lee:2015eva},
we obtain an $\ep$-expansion with high-precision numerical coefficients (up to
$10^4$) near the most relevant dimensions $d=4$ and $d=3$. Using \texttt{PSLQ}
algorithm we recognize these high-precision coefficients in terms of multiple
zeta values (for $d=4-2\ep$) and Euler-Zagier sums (for $d=3-2\ep$). The
obtained results for even and odd dimensions cover all thinkable practical
applications and for the $d=4-2\ep$ case were already successfully applied in
Ref.~\cite{Grozin:2022wse}. The results in $d=3-2\ep$ can find their application
in perturbative calculations in ABJM theory, see,
e.g.~\cite{Bianchi:2017afp,Bianchi:2017ujp}.

\acknowledgments
The work has been supported by Russian Science Foundation under grant
20-12-00205. We are grateful to the Joint Institute for Nuclear Research for
using their supercomputer ``Govorun.''

\appendix

\section{Supplementary files description}
\label{sec:anc-files}

The main results of the article are available in the form of computer readable
files.
For alternating Euler-Zagier sums we use the notation
\begin{equation}
  \label{eq:mzv-def}
  \texttt{mzv[n{\small 1},...,n{\small k}]}  = \zeta_{n_1,\dots,n_k} = \sum\limits_{i_1>\ldots >i_k>0}\prod_{l=1}^{k} \frac{(\operatorname{sign}n_l)^{i_l}}{i_l^{|n_l|}}
\end{equation}

Short description of files and examples of their usage are provided below.

\begin{itemize}

\item[\textsf{HQET4l.st}]~\\
  List of arbitrary $d$ results in the \texttt{SummerTime} package format.
  To
  calculate all integrals in $d=4-2\ep$ with 30 digits precision to the order
  $\mathcal{O}(\ep^2)$ one can run:\\
  \texttt{TriangleSumsSeries[\#,\{ep,2\},30]\& /@ (Get["HQET4l.st"] /.d->4-2*ep)}

\item[\textsf{HQET4l.ldrr}]~\\
  Matrix $L(d)$ of the lowering dimensional recurrence ralation~(\ref{eq:LDRR}).

\item[\textsf{hqetUT4l.3d}]~\\
  List of analytical results for uniform transcendental weight basis unctions expanded near $d=3$ to the
  transcendental weight 10 in terms of alternating Euler-Zagier sums.

\item[\textsf{hqetUT4l.4d}]~\\
  List of analytical results for uniform transcendental weight basis unctions expanded near $d=4$ to the
  transcendental weight 12 in terms of MZV.

\item[\textsf{ut2mi.3d}]~\\
  Conversion matrix from the set of UT basis functions to integrals calculated
  in the present paper(Fig.~\ref{fig:MIs}) for $d=3-2\ep$.  One can obtain list of
  intgrals from basis UT weight functions with:\\
  \texttt{Get["ut2mi.3d"].Get["hqetUT4l.3d"]}

\item[\textsf{ut2mi.4d}]~\\
  Conversion matrix from the set of UT basis functions to integrals calculated
  in the present paper(Fig.~\ref{fig:MIs}) for $d=4-2\ep$.
  One can obtain list of
  intgrals from basis UT weight functions with:\\
  \texttt{Get["ut2mi.4d"].Get["hqetUT4l.4d"]}

\end{itemize}

\bibliographystyle{JHEP}
\bibliography{HQETpropagators}
\end{document}